\documentclass[prl,letterpaper,twocolumn]{revtex4}
\usepackage{graphicx}
\begin{document}
\title{Universal resonant ultracold molecular scattering}
\author{Vladimir Roudnev and Michael Cavagnero}
\affiliation{Department of Physics and Astronomy, University of Kentucky, Lexington, KY 40506-0055}

\begin{abstract}
The elastic scattering amplitudes of indistinguishable, bosonic, strongly-polar molecules possess
universal properties at the coldest temperatures due to wave propagation in the long-range
dipole-dipole field. Universal scattering cross sections and anisotropic threshold angular distributions,
independent of molecular species, result from careful tuning of the dipole moment with an applied electric field. 
Three distinct families of threshold resonances also occur for specific field strengths, and can be both qualitatively and
quantitatively predicted using elementary adiabatic and semi-classical techniques.  
The temperatures and densities of heteronuclear molecular gases required to observe these univeral
characteristics are predicted. 

\end{abstract}
\maketitle

Observations of collective phenemena in BEC's of magnetically dipolar atomic Cr~\cite{Pfau} have spurred attempts to cool
and trap heteronuclear molecules with large electric dipole moments. There has been much recent progress; for example,  
$3\times 10^3$ cm$^{-3}$ OH~\cite{Ye} molecules have been cooled through Stark deceleration and trapped at 30 mK; 
$10^9$ cm$^{-3}$ vibrationally excited 
RbCs~\cite{DeMille} molecules, formed via photoassociation, have been trapped at $250~ \mu K$; and $10^{12}$ cm$^{-3}$ 
(fermionic) KRb~\cite{Jin} molecules in ground ro-vibrational states have been trapped at $350~ nK$.    
A wide variety of techniques are now being developed and applied to produce quantum degenerate gases of such
strongly dipolar molecules, in search of novel collective phenomena and for various applications~\cite{Review1}. 

One might naturally expect that as dilute polar gases are cooled toward the absolute zero of temperature, a regime should 
emerge in which only the longest ranged forces -- dipole-dipole interactions -- are relevant; in which case these different polar 
gases would share a universal equation of state. Myriad theoretical models of ultracold dipolar gases, most based on mean-field
theory, assume purely dipolar interactions between molecules, modified only by a short-ranged species-dependent contact potential.~\cite{Baranov}. 
However, recent detailed calculations of ultracold molecular collisions show strong species-sensitive resonant characteristics of
elastic scattering cross sections, and  
display universality only in the high-temperature regime of semi-classical collision dynamics~\cite{Ticknor1}. 

One purpose of this Letter is to demonstrate several universal characteristics of dipole-dipole collisions that will
emerge at the {\it lowest} temperatures where collective phenomena associated with quantum degeneracy are expected~\cite{Review2}.
Among these universal properties are: 1) an absolute minimum elastic scattering cross-section for zero-energy collisions
of all bosonic dipoles; 2) a strong correlation between threshold angular distributions and the magnitude of threshold
elastic scattering cross sections; 3) several sequences of threshold resonant states that can be tuned with an applied
electric field; and 4) a set of effective barrier heights which retard threshold scattering in all but one elastic channel. 
These features are independent of short-range interactions between the molecular collision partners. 

The near-threshold resonances mentioned above have appeared in earlier calculations
of RbCs and SrO scattering~\cite{TickBohn1,TickBohn2}, and have also recently appeared
in calculations of two molecules in a trap~\cite{Blume}. 
A second purpose of this Letter is to provide a complete classification
and interpretation of these resonances.  

Many molecules possess polar charge distributions in their lowest electronic energy states.
However, in the absence of external fields, as a consequence of parity conservation, even  
strongly ``polar'' molecules have no net dipole moment. 
Strong {\it anisotropic}
molecular interactions therefore require an external electric field, $\vec{\cal E}$, coupling opposite 
parity states, and producing a field-dependent dipole moment  
$\vec \mu({\cal E}) = -(dE/d{\cal E})\hat {\cal E}$
for all molecules~\cite{TickBohn2}. 
The magnitude and sign of the ``induced'' dipole moment, $\mu$, depends on 
details of the $E({\cal E})$ Stark shift of the state of interest for a given molecule. 
Tuning the dipole moment (and therefore the strength of the dipole-dipole interaction) with external field permits variation of
the temperature scale at which effects of quantum degeneracy can be expected. 

At intermolecular separations much greater than the molecular size, molecular interactions reduce to the 
familiar dipole-dipole form
\begin{equation}
  V(\vec r) = \frac{1}{r^3}\left[ \vec\mu_1\cdot\vec\mu_2 - 3(\hat r\cdot\vec\mu_1) (\hat r\cdot \vec\mu_2)\right]
\end{equation}
which, for a pair of identical molecules in the same field-aligned state, yields an 
equation of relative motion
\begin{equation}
  \left[-\frac{\hbar^2}{2M}\nabla^2 +\frac{\mu^2}{r^3}\left( 1 - 3(\hat r\cdot\hat z)^2\right)
 \right]\psi(\vec r) 
                    = E_{\rm rel}\psi(\vec r)
\label{eq:Schro}
\end{equation}
where $\vec r$ is the intermolecular displacement,  $\hat z$ is the field axis and $M$ is the reduced mass.
$\mu$ assumes its maximal value at large fields, but reduces to zero as ${\cal E}\rightarrow 0$. 

Equation (\ref{eq:Schro}) may be written in dimensionless form {\it independent of $M$ and $\mu$} if all distances and energies are measured in dipole units
\begin{equation}
\begin{array}{c}
  D = \mu^2 M/\hbar^2,~~~
  E_D = \hbar^6/(M^3\mu^4)
\end{array}
\label{eq:DipoleScales}
\end{equation}
The emergence of these field-dependent length and energy scales, previously noted by many authors~\cite{Scale}, is striking when one recognizes that 
{\it for maximal} $\mu$, $D\sim 10^2 - 10^6$ times {\it larger} than typical molecular length scales, while 
$E_D\sim 10^{-4} - 10^{-12}$ times {\it smaller} than typical rotational level splittings. 
For the examples cited in our introduction, we note
that $D\sim 30$ a.u. and $E_D\sim 10^{-2}K$ for both triplet KRb and atomic Cr; $D$ is of order $10^{4}$ a.u. for OH, singlet
KRb, and ground state RbCs, while $E_D\sim 1-100$ nK. For a stronger alkali-halide dipole such as LiF, 
$D\sim 10^6$ a.u. and $E_D\sim 0.1$ nK. Note that the recent singlet (fermionic) KRb experiment is within a factor of 4 in temperature
of the intrinsic dipole energy scale~\cite{Jin}, and the gas density is comparable to $1/D^3$. Note also
that the energy scales (and relevant temperatures) cited here can be increased by {\it decreasing} the applied electric field. 

There are two restrictions on the domain of validity of Eq. (\ref{eq:Schro}). 
The dipole-dipole interaction accurately represents the intermolecular potential only where 
higher multipoles and exchange terms are negligible. Within an exchange radius $\sim 20-30$ Bohr, 
intermolecular interactions are complex and species-dependent. There is also a range of separations 
at which the fields created by the dipoles dominate the external field, so that effects of collisional 
depolarization become important for $r\sim (d/{\cal E})^{1/3}$, typically $\sim$ 100 Bohr in moderate fields 
for strongly polar molecules. Key to the present study is that both of these length scales are orders 
of magnitude smaller than $D$ for many systems. 

\begin{figure}[htb]
\begin{center}
	\includegraphics[scale=0.27,clip=true]{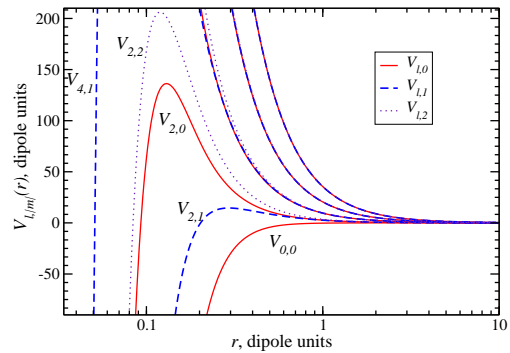}
\end{center}
\caption{\label{Fig:1} (Color online) Adiabatic potential curves for bosonic polar molecules. The curves are labeled by their asymptotic
angular momentum $\ell$ and by the magnitude of the conserved magnetic quantum number $m$. } 
\end{figure}

Equation~\ref{eq:Schro} is usually analyzed by partial wave expansion~\cite{Ticknor2}. 
In contrast, we adopt here Born-Oppenheimer adiabatic states, 
whose eigenenergies relevant to bosonic molecules are displayed in dipolar units in Fig.~\ref{Fig:1}. 
The potential curves and coupling matrix elements of this representation are {\it universal} (i.e. independent of 
collision partner, reduced mass, dipole moment, and electric field strength) for all $r$ at which the dipole-dipole interaction
is dominant. 
Fig.~\ref{Fig:1} shows that an infinity of channels (corresponding to the infinity of partial waves) produce 
a single attractive potential, $V_{0,0}$, and an infinite number of potentials with centrifugal barriers. 

%%%%%%%%%%%%%%%%%%%
%%%%%%%%%%%%%%%%%%%

Due to cylindrical symmetry about the field axis, the adiabatic states couple
only to other states of the same $m$, which is strictly conserved.
The adiabatic curves approach familiar centrifugal potentials at large $r$, permitting them to be labeled
by their large-$r$ angular momentum quantum number, $\ell$, as well. Accordingly, the curves in Fig.~\ref{Fig:1},
and the resonant states they support, are uniquely designated by $(\ell,\vert m\vert)$.   
Note that $V_{2,1}$ and $V_{2,0}$ present universal barriers of heights $\sim 14~E_D$ and $\sim 140~E_D$, respectively. 
Since $m$ is conserved, the $V_{2,1}$ channels are not coupled to the $V_{0,0}$ channel, while $V_{2,0}$
and the ground state channel are strongly coupled. Accordingly, at energies much 
smaller than $\sim 140~E_D$, scattering is dominated by the two lowest adiabatic potentials, 
absent the resonances discussed below. 

The most salient feature of Fig. (1) is that {\it the barrier heights are sufficiently high that tunneling is
strongly suppressed in all channels near threshold.} Accordingly, molecules with energies $<< 14~E_D$ 
can approach one another to distances 
less than $0.2 D$ only through the attractive $V_{0,0}$ potential well. While strong non-adiabatic coupling
exists both inside and outside the potential barriers, the two regions are effectively decoupled by tunneling 
suppression. It is this feature that leads to universal scattering characteristics that are 
independent of short-range physics. This independence is illustrated here by applying a variable hard-sphere boundary
condition at a small radius, $r_0$.

We have solved the coupled-channel Schr\"odinger equation in the adiabatic representation numerically~\cite{DeBooorSchwartz}, 
propagating from $r_0$ to asymptotic distances $r\sim 500 D$. 
While non-adiabatic coupling cannot be neglected, the adiabatic basis converges far more rapidly than a partial
wave expansion near threshold~\cite{RC2}.

\begin{figure}[htb]
\begin{center}
\includegraphics[scale=0.27,clip=true]{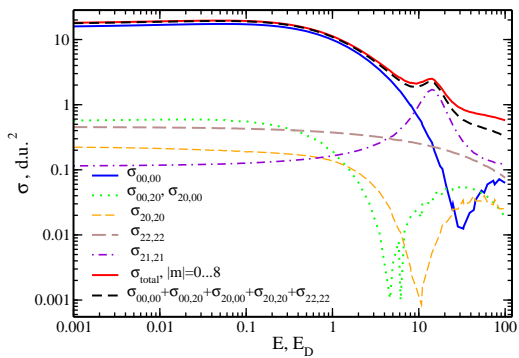}
\caption{\label{Fig:2} (Color online) Channel contributions to the threshold cross section for $r0=0.0111$ dipole units. The $(0,0)$ and degenerate $(2,1)$ channels dominate scattering below $\sim 14~E_D$.}
\end{center}
\end{figure}

\begin{figure}[htb]
\begin{center}
	\includegraphics[scale=0.27, clip=true]{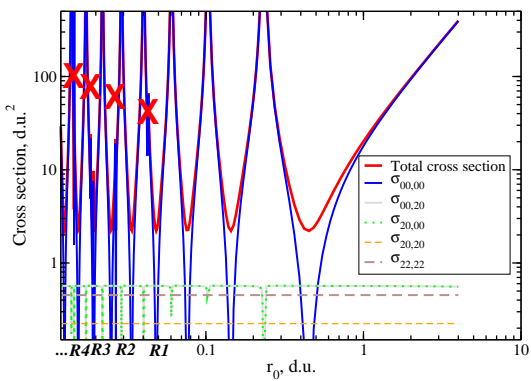}
\caption{\label{Fig:3} (Color online) Near-threshold cross section as a function of the cut-off radius. Positions of the $(2,0)$ Feshbach-shape resonances are marked by {\bf X}s and labeled as {\em R1, R2, R3, R4}.}
\end{center}
\end{figure}

The contribution of various channels to the total elastic cross-section for bosonic scattering (averaged over the electric-field direction), is shown in
Fig.~\ref{Fig:2}, for $r_0 = 0.0111$ dipole units. 
The uncoupled $V_{0,0}$ and $V_{2,1}$ channels dominate scattering below $\sim 14E_D$, as indicated above.  
Note, however, that detailed convergence of the cross-section (even at energies less than a nano-Kelvin) requires several adiabatic channels. 
Both the magnitude of the threshold cross section and the degree of convergence vary with the hard-sphere radius, as 
they are sensitive to resonance formation in the inner wells of the potential curves. 
This is illustrated in Fig.~\ref{Fig:3}, where the elastic cross section (averaged over the field direction) is plotted versus hard-sphere radius at the near-threshold 
energy, $E = 10^{-3}~E_D$. 
Note that the threshold cross-section varies by 4-orders of magnitude, depending on the specific details  of the short-range scattering. 

%%%%%%%%%%%%%%%%%%%
%%%%%%%%%%%%%%%%%%%

Figure~\ref{Fig:3} also illustrates the partial wave content of the scattering cross sections. 
The near-threshold cross section varies quasi-periodically with $r_0$ due to successive ``Ramsauer minima'' in the attractive adiabatic channel (i.e. varying $r_0$ alters the number of bound states in the ground state potential curve, and correspondingly sweeps the s-wave scattering length through complete cycles). 
Despite this sensitivity, the cross section receives a contribution of $2D^2$ {\it independent of $r_0$} from all asymptotically repulsive channels. 
Accordingly, at a minimum of the total cross section, the $(0,0)$ channel is effectively ``turned off'', and the
long-range part of the intermolecular interaction dominates. 
Absent the resonances, the threshold cross section is accordingly governed purely by the long-range physics of the
dipole-dipole interaction~\cite{Threshold}. 

%%%%%%%%%%%%%%%%%%%%%%%%
%%%%%%%%%%%%%%%%%%%%%%%%

The resonances shown in Fig. 3 can be predicted by semiclassical quantization in accurate fits to
the potential curves~\cite{RC2}. The positions of the successive Ramsauer peaks in the $(0,0)$ potential 
are given (to 3 significant figures) by 
\begin{equation}
 	r_{0}^{0,0}(n)=\frac{1}{4.37+3.46n+0.675n^2} \ \ (d.u.) .
 	\label{eq:sWaveRes}
\end{equation}
Similarly, the semiclassical description predicts positions of resonances in the $V_{2,0}$ potential that are marked as $R1$, $R2$,
$R3$ etc. in Fig.~\ref{Fig:3}, and given by 
\begin{equation}
 	r_{0}^{2,0}(n)=\frac{1}{10.88+11.81 n+ 0.685 n^2} \ \ (d.u.) .
 	\label{eq:dWaveRes}
\end{equation}
If nonadiabatic coupling were negligible, the $(2,0)$ resonances would have a near-zero (tunneling) width when approaching threshold
because of the large dipole-modified centrifugal barrier in Fig. 1. To the extent that tunneling is negligible, these are effectively
discrete states coupled to the ground state continuum. 
Their actual width is due to decay into the lowest adiabatic channel through short-range nonadiabatic coupling. Accordingly, the $(2,0)$ resonances 
appear as Feshbach/Fano resonances in the total elastic cross section. 
Finally, there is also a series of extremely narrow tunneling resonances in the $V_{2,1}$ potentials, whose positions are given by 
\begin{equation}
 	r_{0}^{2,1}(n)=\frac{1}{3.975+7.973n+ 0.675n^2} \ \ (d.u.) .
 	\label{eq:TunRes}
\end{equation}
but these have little relevance to threshold elastic scattering, as the $\vert m\vert =1$ channels are not coupled to
the ground state channel. 

\begin{figure*}[htb]
\begin{center}
  \begin{tabular}{cccc}
  %% row 1
   (a) $r_0=0.014013$~d.u.
  &  
   (b) $r_0=0.014549$~d.u.
  &
   (c) $r_0=0.014576$~d.u.
  &
   (d) $r_0=0.014579$~d.u.
\\ 
 	\includegraphics[scale=0.27, clip=true]{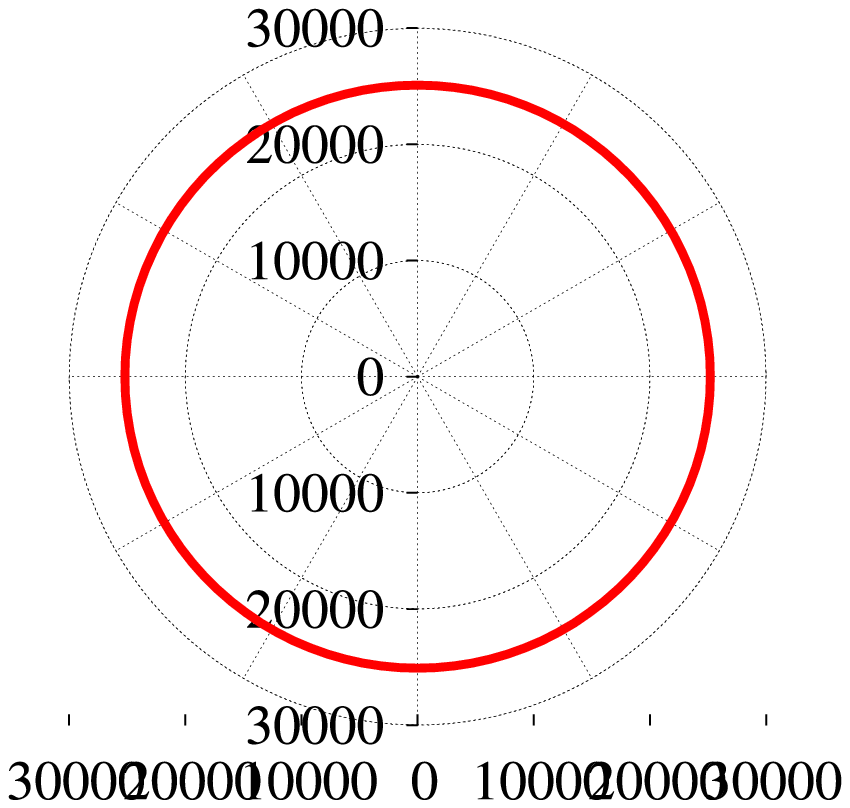}
  &
  	\includegraphics[scale=0.27, clip=true]{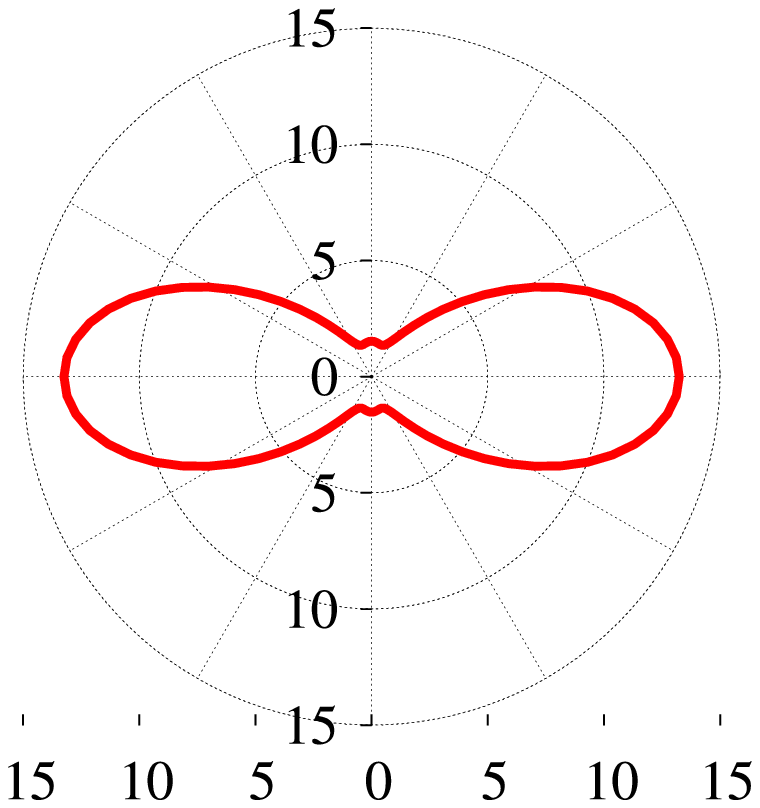}
  &
  	\includegraphics[scale=0.27, clip=true]{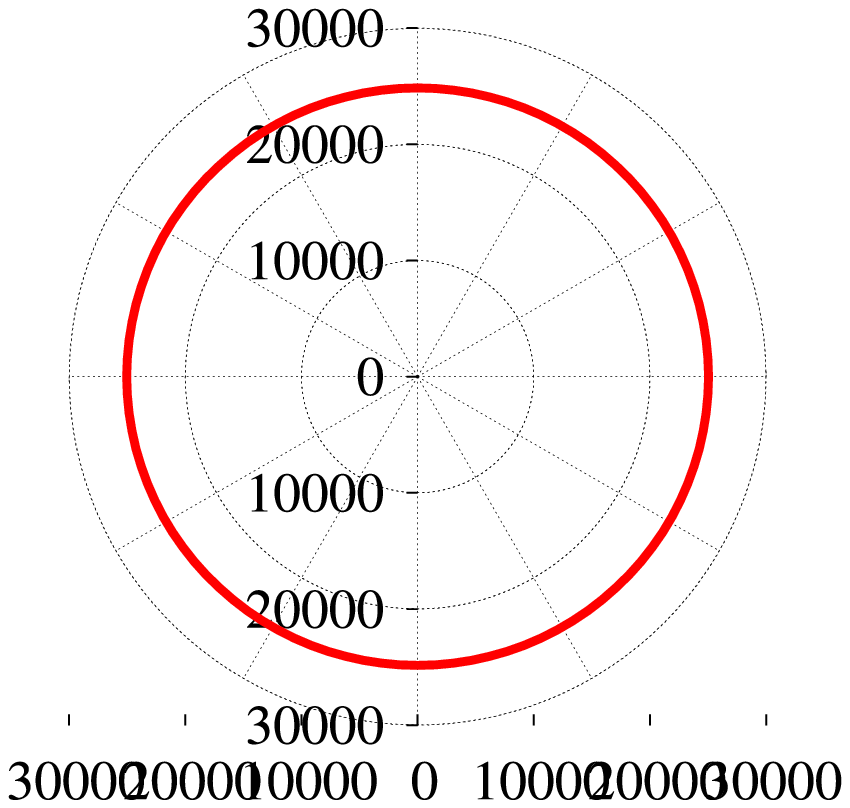}
  &
  	\includegraphics[scale=0.27, clip=true]{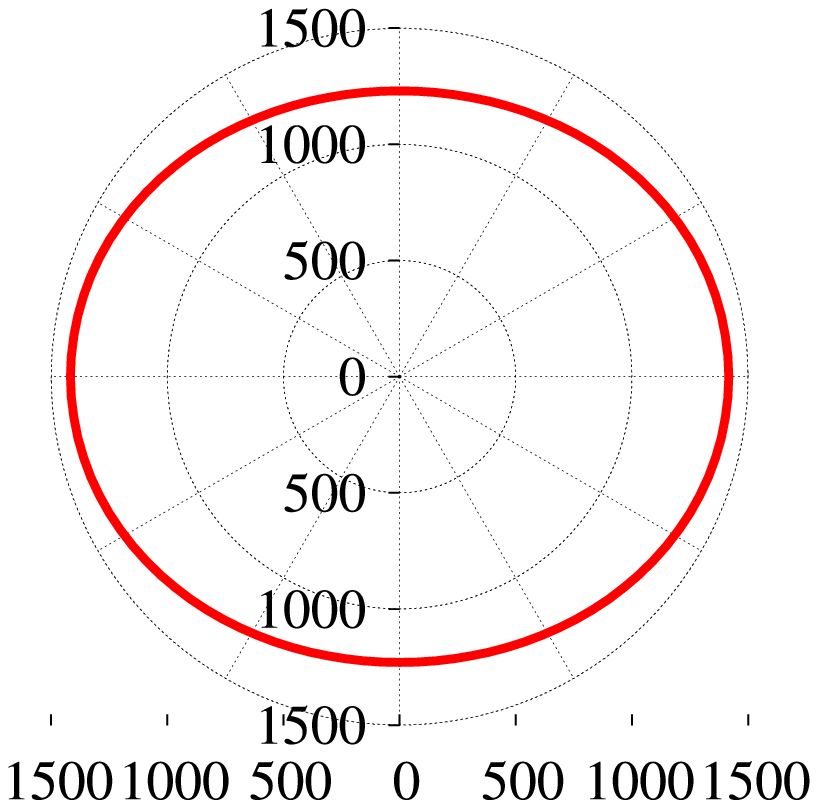}
  \end{tabular}
\caption{\label{Fig:Anisotropy} (Color online) Cross section dependence on the angle between the incident direction and the field axis, for
	values of the cut off radius in the vicinity of the Feshbach/shape resonance. 
         When the total  cross section is large, the scattering is almost isotropic, even at the ``d-wave''	resonance, (c). 
         }
\end{center}
\end{figure*}

The dual Feshbach/shape nature of the $(2,0)$ resonance leads to strong correlation between the anisotropy of the differential cross section and the 
magnitude of the total cross section. 
One might expect strongly anisotropic scattering  
near a $(2,0)$ resonance. 
However, due to the small value of the tunneling amplitude, 
the inner well of the $(2,0)$ potential is only accessed through coupling to
the $(0,0)$ channel. 
As illustrated in Fig. 4, anisotropy results only when the $(0,0)$ channel is ``turned off'' (either near a Ramsauer minimum
or -- due to interference -- to the side of a Feshback/shape resonance), 
and stems from long-range non-adiabatic coupling {\it outside} the $(2,0)$ potential barrier, as well as from 
the uncoupled $(2,2)$ and $(2,1)$ channels. 
As a result, whenever the scattering cross section peaks, the scattering becomes isotropic, even at the peak of the $(2,0)$ resonance;
while near the minimum of the total cross section, a strongly anistropic distribution emerges. 
This view augments the interpretation of similar resonances in atomic collisions in very strong fields~\cite{Mari}. 
%%%%%%%%%%%%%%%%%%%
\begin{figure}[htb]
\begin{center}
	\includegraphics[scale=0.27, clip=true]{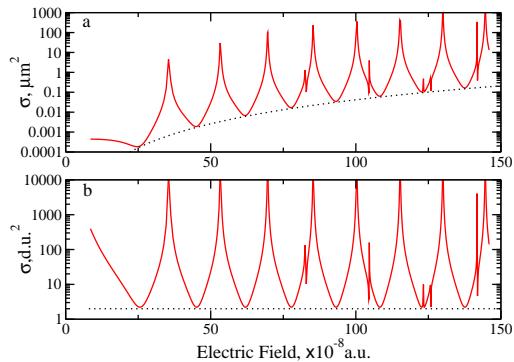}
\caption{\label{fig:FieldDependence} Electric field dependence of the near-threshold LiF elastic scattering cross section, in both (a) absolute and (b) dipole units. 
The minima of the cross section ($=2D^2$) in the upper figure rise rapidly with the field as a result of rapidly growing dipole length.}
\end{center}
\end{figure}

While the existence of threshold resonances depends on $r_0$ and will, accordingly, vary from molecule to molecule, the
resonances can be observed (or eliminated!) for any given molecular species by varying the applied field~\cite{TickBohn2}. 
We described above a dimensionless threshold cross section that varies with a dimensionless cut-off $\sigma_0[r_0 (d.u.)]$. 
To the extent that short-range physics is insensitve to the external field, the observed elastic scattering cross section will scale with the field ${\cal E}$ as 
$\sigma =D({\cal E})^2 \sigma_0[r_0(a.u.)/D({\cal E})]$, 
where $D({\cal E})$ is given by Eq.~\ref{eq:DipoleScales}. 
This field dependence is displayed in Fig.~\ref{fig:FieldDependence} for the LiF molecule, with $r_0$ chosen arbitrarily at $80$ a.u.,
and assuming a quadratic field variation of the LiF ground state. 
The actual, physical value of $r_0$ can be determined empirically by observation of the first peak at the smallest
value of the applied electric field. This, in turn, lends predictive value to Eq.~(\ref{eq:sWaveRes}), which, with knowledge of
$\mu({\cal E})$, predicts the field
values for all subsequent $(0,0)$ resonances. 
The cross-section is displayed in both conventional and in dipole units. 
Also shown in the figure is the minimal value of $2D^2$ and its variation with field strength. 

The adiabatic potential curves serve to identify $E<14 E_D$ as the universal energy range of ultracold 
molecular collisions, and $\sigma \sim 2 D^2$ as the {\it minimum} elastic cross section for threshold collisions between polarized
molecules. Far greater
cross sections approaching the unitarity limit $\sim 4\pi D^2/(E/E_D)$ are achievable with small variations of the applied field: 
note the micron$^2$ scale in Fig.~\ref{fig:FieldDependence}(a).  
Observation of the universal properties predicted here requires a stongly dipolar gas whose temperature is $\sim E_D/k$ and
whose density is less than $1/D^3$ such that two-body collisions dominate. Recent experimental efforts are approaching
this regime. 

The authors are grateful to John Bohn for helpful conversations and a critical reading of the manuscript.

\end{document}